\documentclass[conference]{IEEEtran}
\IEEEoverridecommandlockouts
\usepackage{fancyhdr}
\fancypagestyle{cfooter}{ %
\fancyhf{} 
\cfoot{\scriptsize{This work has been submitted to the IEEE for possible publication. Copyright may be transferred without notice, after which this version may no longer be accessible.}
}

}

\usepackage{cite}
\usepackage{amsmath,amssymb,amsfonts}
\usepackage{algorithmic}
\usepackage{graphicx}
\usepackage{textcomp}
\usepackage{xcolor}
\def\BibTeX{{\rm B\kern-.05em{\sc i\kern-.025em b}\kern-.08em
    T\kern-.1667em\lower.7ex\hbox{E}\kern-.125emX}}

\usepackage{cite}
\usepackage{amsmath,amssymb,amsthm,fixmath}
\usepackage{mathtools}
\usepackage{xcolor}
\usepackage{bm}
\usepackage{makecell}
\usepackage{pgfplots}
\usepackage{fancyhdr}
\usepackage{hyperref}
\usepackage{caption}
\usepackage{xcolor}
\usepackage{bm}
\usepackage{makecell}
\usepackage{pgfplots}
\usepackage{tikz}
\usepackage{graphicx}
\usepackage{booktabs} 
\usetikzlibrary{intersections,pgfplots.fillbetween,patterns,spy,shapes.geometric,calc,positioning}
\usetikzlibrary{arrows,decorations.pathmorphing,backgrounds,fit,matrix,fillbetween}
\usetikzlibrary{shapes.geometric, arrows.meta}
\usetikzlibrary{decorations.pathreplacing,decorations.markings}
\usepgfplotslibrary{groupplots}
\usepackage{soul}
\usepackage{url}
\usepackage{float}
\usepackage{adjustbox}
\usetikzlibrary {arrows.meta}
\usepackage{acro}
\usepackage{balance}

\DeclareMathOperator{\E}{\mathbb{E}}

\DeclareAcronym{CS}{short = CS ,long = compressive sensing}
\DeclareAcronym{MMSE}{short = MMSE,long =  minimum mean squared error}
\DeclareAcronym{MSE}{short = MSE, long = mean squared error}
\DeclareAcronym{NMSE}{short = NMSE, long = normalized MSE}
\DeclareAcronym{LMMSE}{short = LMMSE, long = linear minimum mean squared error}
\DeclareAcronym{WSS}{short = WSS, long = wide sense stationary}
\DeclareAcronym{DFT}{short = DFT, long = discrete Fourier transform}
\DeclareAcronym{FT}{short = FT, long = Fourier transform}
\DeclareAcronym{EM}{short = EM, long = expectation-maximization}
\DeclareAcronym{BIC}{short = BIC, long = Bayesian information criterion}
\DeclareAcronym{AIC}{short = AIC, long = Akaike information criterion}
\DeclareAcronym{i.i.d.}{short = i.i.d., long = independent and identically distributed}
\DeclareAcronym{FLOP}{short = FLOP, long = floating point operation}
\DeclareAcronym{AWGN}{short = AWGN, long = additive white Gaussian noise}
\DeclareAcronym{LS}{short = LS, long = least squares}
\DeclareAcronym{ARMA}{short = ARMA, long = autoregressive moving-average}
\DeclareAcronym{FBM}{short = FBM, long = fractional Brownian motion}
\DeclareAcronym{FFT}{short = FFT, long = fast Fourier transform}
\DeclareAcronym{GMM}{short = GMM, long = Gaussian mixture model}
\DeclareAcronym{VAE}{short = VAE, long = variational autoencoder}
\DeclareAcronym{NN}{short = NN, long = neural network}
\DeclareAcronym{PD}{short = PD, long = positive definite}
\DeclareAcronym{OP}{short = OP, long = optimization problem}
\DeclareAcronym{GS}{short = GS, long = Gohberg-Semencul}
\DeclareAcronym{CSI}{short = CSI, long = channel state information}
\DeclareAcronym{SAVG}{short = SAVG, long = SCM averaged along its diagonals}
\DeclareAcronym{UE}{short = UE, long = user equipment}
\DeclareAcronym{DoA}{short = DoA, long = direction of arrival}
\DeclareAcronym{DoAs}{short = DoAs, long = directions of arrival}
\DeclareAcronym{DoDs}{short = DoDs, long = directions of depature}
\DeclareAcronym{ML}{short = ML, long = machine learning}
\DeclareAcronym{BS}{short = BS, long = base station}
\DeclareAcronym{WSSUS}{short = WSSUS, long = wide-sense-stationary-uncorrelated-scattering}
\DeclareAcronym{ULA}{short = ULA, long = uniform linear array}
\DeclareAcronym{URA}{short = URA, long = uniform rectangular array}
\DeclareAcronym{MIMO}{short = MIMO, long = multiple-input-multiple-output}
\DeclareAcronym{SIMO}{short = SIMO, long = single-input-multiple-output}
\DeclareAcronym{OFDM}{short = OFDM, long = orthogonal-frequency-division-multiplexing}
\DeclareAcronym{mmWave}{short = mmWave,long =  millimeter wave}
\DeclareAcronym{HST}{short = HST, long = high-speed train}
\DeclareAcronym{UAV}{short = UAV, long = unmanned aerial vehicles}
\DeclareAcronym{IoT}{short = IoT, long = Internet of things}
\DeclareAcronym{Eig}{short = Eig, long = eigenvalue}
\DeclareAcronym{Frob}{short = Frob, long = Frobenius}
\DeclareAcronym{PGD}{short = PGD, long = projected gradient descent}
\DeclareAcronym{PLS}{short = PLS, long = projected LS}
\DeclareAcronym{LOS}{short = LOS,  long =line-of-sight}
\DeclareAcronym{GAN}{short = GAN, long = generative adversarial network}
\DeclareAcronym{NLOS}{short = NLOS, long = non-line-of-sight}
\DeclareAcronym{SISO}{short = SISO, long = single-input-single-output}
\DeclareAcronym{SNR}{short = SNR, long = signal-to-noise ratio}
\DeclareAcronym{BN}{short = BN, long = Bayesian network}
\DeclareAcronym{DM}{short = DM, long = Diffusion model}
\DeclareAcronym{KL}{short = KL, long = Kullback-Leibler}
\DeclareAcronym{ELBO}{short = ELBO, long = evidence lower bound}
\DeclareAcronym{CGLM}{short = CGLM, long = conditionally Gaussian latent model}
\DeclareAcronym{PGMM}{short = PGMM, long = parameter GMM}
\DeclareAcronym{PVAE}{short = PVAE, long = parameter VAE}
\DeclareAcronym{AP}{short = AP, long = access point}

\DeclareAcronym{CDL}{short = CDL,long = cluster delay line}
\DeclareAcronym{TDL}{short = TDL,long = tap delay line}
\DeclareAcronym{EPA}{short = EPA,long = extended pedestrian A}
\DeclareAcronym{GSCM}{short = GSCM,long = geometry-based stochastic channel model}
\DeclareAcronym{CME}{short = CME,long = conditional mean estimator}

\DeclareAcronym{CSGMM}{short = CSGMM,long = compressive sensing \ac{GMM}}
\DeclareAcronym{CSVAE}{short = CSVAE,long = compressive sensing \ac{VAE}}
\DeclareAcronym{CP-GMM}{short = CP-GMM,long = channel parameter-\ac{GMM}}
\DeclareAcronym{CP-VAE}{short = CP-VAE,long = channel parameter-\ac{VAE}}
\DeclareAcronym{SBL}{short = SBL,long = sparse Bayesian learning}
\DeclareAcronym{SBGM}{short = SBGM, long  = sparse Bayesian generative modeling}
\DeclareAcronym{PG}{short = PG, long  = path gain}
\DeclareAcronym{GM}{short = GM, long  = Generative model}
\DeclareAcronym{DL}{short = DL, long  = deep learning}
\DeclareAcronym{LLM}{short = LLM, long  = large language model}
\DeclareAcronym{JCAS}{short = JCAS, long  = joint communication and sensing}
\DeclareAcronym{RMSE}{short = RMSE, long  = root mean squared error}

\definecolor{color1}{rgb}{0 0.4470 0.7410}
\definecolor{color2}{rgb}{0.3010 0.7450 0.9330}
\definecolor{color3}{rgb}{0.4940 0.1840 0.5560}
\definecolor{color4}{rgb}{0.6350 0.0780 0.1840}
\definecolor{color5}{rgb}{0.8500 0.3250 0.0980}
\definecolor{color6}{rgb}{0.9290 0.6940 0.1250}
\definecolor{color7}{rgb}{0 0.50 0}
\definecolor{color8}{rgb}{0.4660 0.6740 0.1880}
	
\definecolor{Yellow}{rgb}{1.00, 0.71, 0.00}
\definecolor{Orange}{rgb}{1.00, 0.50, 0.00}
\definecolor{Red}{rgb}{0.90, 0.20, 0.09}
\definecolor{DarkRed}{rgb}{0.9, 0.3, 0.3}
\definecolor{Blue}{rgb}{0.00, 0.60, 1.00}
\definecolor{LightBlue}{rgb}{0.25, 0.75, 1.00}
\definecolor{Green}{rgb}{0.57, 0.67, 0.42}
\definecolor{LightGreen}{rgb}{0.71, 0.89, 0.58}
\definecolor{Black}{rgb}{0,0,0}
\definecolor{Gray}{rgb}{0.7,0.7,0.6}
\definecolor{DarkGray}{rgb}{0.4,0.4,0.3}
\definecolor{color1bg}{HTML}{ECD9ED}
	
\definecolor{MidnightBlue}{HTML}{006795}
\definecolor{SpringGreen}{HTML}{C6DC67}
\definecolor{PineGreen}{HTML}{008B72}
\definecolor{Maroon}{HTML}{AF3235}
\definecolor{RedOrange}{HTML}{F26035}
\definecolor{SkyBlue}{HTML}{46C5DD}
\definecolor{Dandelion}{HTML}{F7921D}
\definecolor{Periwinkle}{HTML}{7977B8}
\definecolor{RedViolet}{HTML}{3C8031}

\definecolor{asparagus}{rgb}{0.53, 0.66, 0.42}
\definecolor{burntsienna}{rgb}{0.91, 0.45, 0.32}
\definecolor{cadetblue}{rgb}{0.37, 0.62, 0.63}
\definecolor{carminepink}{rgb}{0.92, 0.3, 0.26}
\usetikzlibrary{external}
\begin{document}
\title{Sparse Bayesian Generative Modeling for \\ Joint Parameter and Channel Estimation\\
\thanks{
This work was supported by the Federal Ministry of
Education and Research of Germany.
Joint project 6G-life, project ID: 16KISK002.
}
}

\author{\IEEEauthorblockN{Benedikt Böck, Franz Weißer, Michael Baur, and Wolfgang Utschick}
\IEEEauthorblockA{\textit{TUM School of Computation, Information and Technology, Technical University of Munich, Germany}}
Email: \{benedikt.boeck, franz.weisser, mi.baur, utschick\}.tum.de}

\maketitle

\thispagestyle{cfooter}

\begin{abstract}
Leveraging the inherent connection between sensing systems and wireless communications can improve their overall performance and is the core objective of joint communications and sensing. For effective communications, one has to frequently estimate the channel. Sensing, on the other hand, infers properties of the environment mostly based on estimated physical channel parameters, such as directions of arrival or delays. 
This work presents a low-complexity generative modeling approach that simultaneously estimates the wireless channel and its physical parameters without additional computational overhead. To this end, we leverage a recently proposed physics-informed generative model for wireless channels based on sparse Bayesian generative modeling and exploit the feature of conditionally Gaussian generative models to approximate the conditional mean estimator.
\end{abstract}


\section{Introduction}
\acp{GM} play a fundamental role in today's rapid advancements of \ac{DL} \cite{Bond2022}. Most research focuses on \acp{LLM} for text generation \cite{zhao2024} and image generation models \cite{rombach2022}, but also in wireless communication and, especially, wireless channel modeling, \acp{GM} promise performance gains \cite{Yang2019}. However, compared to text and image generation, \acp{GM} for channel modeling are subject to very different restrictions and conditions. Unlike text and natural images, wireless channels follow the fundamental laws of electromagnetic wave propagation. Moreover, when using the \ac{GM} during online operation, its computational complexity is more important than in standard text and image applications. Offloading the \ac{GM} to mobile devices constrains the number of \ac{GM} parameters. In addition, acquiring high-quality training data for the \acp{GM} is difficult and potentially requires costly measurement campaigns.

To overcome these issues, the work in \cite{boeck24_icml} builds upon \ac{SBGM} \cite{boeck2024nips} and introduces a physics-informed \ac{GM} for wireless channels, which simultaneously can learn from noisy channel observations, generalizes to arbitrary system configurations and obeys the underlying physics. This model is evaluated in \cite{boeck24_icml} regarding its generation capability for, e.g., providing training data. 

In our work, we show that the model from \cite{boeck24_icml} can also be directly used for improving physical layer applications and, particularly, parameter and channel estimation. Specifically, \acp{CGLM}, i.e., \acp{GM} with a Gaussian conditioned on a latent variable, such as \acp{GMM} and \acp{VAE} have been used for channel estimation in \cite{Koller22_ce, baur24_ce, Fesl24_ce, boeck23_ce,Fesl2022}. Since \ac{SBGM} involves a \ac{CGLM} \cite{boeck2024nips}, the model in \cite{boeck24_icml} can be used in the same way. 

Our main contributions are as follows. First, we show that the \ac{GMM}-based estimators from \cite{Fesl2022}, which use a \ac{DFT}-based parameterization of the \ac{GMM} covariances, are a special case of the model in \cite{boeck24_icml}. Based on this reinterpretation, we show how the model in \cite{boeck24_icml} can be used to jointly estimate parameters (e.g., \ac{DoAs} and delays) and the channel without computational overhead, rendering it particularly interesting for applications in \ac{JCAS}. In addition, by leveraging the integrated physics, we reduce our estimator's complexity to be linear in the channel dimension (e.g., the number of antennas). Moreover, the number of model parameters does not scale with the channel dimension and is in the range of a few hundred while still achieving state-of-the-art performance.

\section{Preliminaries}

\subsection{Gaussian Mixture Models for Channel Estimation}
\acp{GMM} are \acp{GM}, which model the unknown distribution $p(\bm{h})$ of interest as a mixture of $K$ Gaussians, i.e.,
\begin{equation}
    \label{eq:GMM_dist}
    p(\bm{h}) = \sum_{k=1}^K \rho_k \mathcal{N}_\mathbb{C}(\bm{h};\bm{\mu}_k,\bm{C}_k)
\end{equation}
with the (prior) weights $\rho_k$, the means $\bm{\mu}_k$ and the covariances $\bm{C}_k$ being the model parameters. Learning these parameters based on a ground-truth dataset $\mathcal{H} = \{\bm{h}^{(n)}\}_{n=1}^{N_t}$ can be done by an \ac{EM} algorithm \cite{Dempster1977, Bishop2007}. While not incorporating \acp{NN} and, thus, not belonging to the \ac{DL}-subfield of \ac{ML}, \acp{GMM} offer several benefits for wireless communications. First, they can be extended to learn from noisy observations \cite{Yang2015,Fesl2023compressed}. Second and arguably more important, \acp{GMM} enable a closed-form computation of the \ac{CME} \cite{Yang2015}, which minimizes the \ac{MSE} and, thus, is a desired channel estimator, i.e.,
\begin{equation}
    \label{eq:gmm_cme}
    \E[\bm{h}|\bm{y}] = \E[\E[\bm{h}|\bm{y},k]|\bm{y}] = \sum_{k=1}^K p(k|\bm{y})\bm{\mu}_k^{\bm{h}|\bm{y},k}
\end{equation}
with $\bm{y} = \bm{A}\bm{h} + \bm{n}$, measurement matrix $\bm{A}$ and noise realization
$\bm{n} \sim \mathcal{N}_\mathbb{C}(\bm{0},\sigma_n^2\operatorname{\mathbf{I}})$. Moreover \cite{Koller22_ce}, 
\begin{align}
    \bm{\mu}_k^{\bm{h}|\bm{y},k} = \bm{C}_k\bm{A}^{\operatorname{H}}(\bm{A}\bm{C}_k\bm{A}^{\operatorname{H}} + \sigma_n^2\operatorname{\mathbf{I}})^{-1}(\bm{y} - \bm{A}\bm{\mu}_k) + \bm{\mu}_k,\\
    p(k|\bm{y}) = \frac{\rho_k \mathcal{N}_\mathbb{C}(\bm{y};\bm{A}\bm{\mu}_k,\bm{A}\bm{C}_k\bm{A}^{\operatorname{H}} + \sigma_n^2\operatorname{\mathbf{I}})}{\sum_{i}\rho_i \mathcal{N}_\mathbb{C}(\bm{y};\bm{A}\bm{\mu}_i,\bm{A}\bm{C}_i\bm{A}^{\operatorname{H}} + \sigma_n^2\operatorname{\mathbf{I}})}.
\end{align}
\subsection{Utilizing Structured Covariances in GMMs}
To reduce the number of model parameters and online complexity, the work in \cite{Fesl2022} investigates \ac{DFT}-based parameterizations of $\bm{C}_k$ in \eqref{eq:GMM_dist}. Specifically, $\bm{C}_k$ is decomposed as
\begin{equation}
    \label{eq:C_decom}
    \bm{C}_k = \bm{Q}\text{diag}(\bm{c}_k)\bm{Q}^{\operatorname{H}}
\end{equation}
with $\bm{Q}$ being either a squared $N \times N$, or a $N \times 2N$-over-sampled \ac{DFT} matrix. Former leads to $\bm{C}_k$ being circulant, while latter leads to $\bm{C}_k$ being Toeplitz.\footnote{The work in \cite{Fesl2022} uses $\bm{Q}^{\operatorname{H}}$ first in their decomposition, which covers the same set of matrices as \eqref{eq:C_decom} and, thus, is equivalent.} Moreover, the work in \cite{boeck2024wcl} shows that \acp{GMM} learn their means $\bm{\mu}_k$ in \eqref{eq:GMM_dist} to be zero when being properly trained on wireless channels. Thus, the means $\bm{\mu}_k$ can also be constrained to be zero to further reduce the number of model parameters.

\subsection{Sparse Bayesian Generative Models for Wireless Channels}
The work in \cite{boeck2024nips} introduces \ac{SBGM}, which combines \ac{SBL} \cite{Tipping2001,Wipf2004} with \acp{CGLM} resulting in a simultaneously sparsity-inducing and learnable statistical model. Next to a \ac{VAE}-based variant, \cite{boeck2024nips} introduces a \ac{GMM}-based implementation of \ac{SBGM}, the \ac{CSGMM}, whose statistical model is given by
\begin{align}
\label{eq:y_s}
    \bm{y}|\bm{s} \sim p(\bm{y}|\bm{s}) = \mathcal{N}_\mathbb{C}(\bm{y};\bm{A}\bm{D}\bm{s},\sigma^2\operatorname{\mathbf{I}}),\\
    \label{eq:s_z}
    \bm{s}|k \sim p(\bm{s}|k) = \mathcal{N}_\mathbb{C}(\bm{s};\bm{0},\text{diag}(\bm{\gamma}_k)),\\ 
    \label{eq:z}
    k \sim p(k) = \rho_k
\end{align}
with $\bm{y} = \bm{A}\bm{x} + \bm{n}$ ($\bm{n} \sim \mathcal{N}_\mathbb{C}(\bm{0},\sigma^2 \operatorname{\mathbf{I}})$) being a noisy and compressed observation of some signal $\bm{x}$ of interest. The signal $\bm{x}$ is assumed to be compressible with respect to some pre-known dictionary $\bm{D}$ with compressible representation $\bm{s}$, i.e., $\bm{x} = \bm{D}\bm{s}$. For training, \cite{boeck2024nips} derives an extended \ac{EM} algorithm which allows to learn the model parameters $\{\bm{\gamma}_k,\rho_k\}_{k=1}^K$ in \eqref{eq:y_s}-\eqref{eq:z} solely based on a training dataset of noisy and compressed observations $\mathcal{Y} = \{\bm{y}^{(n)}\}_{n=1}^{N_t}$.

In \cite{boeck24_icml}, the statistical model \eqref{eq:y_s}-\eqref{eq:z} is applied to modeling wireless channels by associating the signal $\bm{x}$ with the channel $\bm{h}$. While \cite{boeck24_icml} considers spatial, as well as time-varying and wideband channels, we only discuss spatial channels in this work. However, our findings can be directly extended to the other cases. To incorporate the underlying physics, \cite{boeck24_icml} utilizes that wireless channels are compressible with respect to a physics-related dictionary, which is typically utilized for parameter estimation (cf. \cite{Dai2021}). Specifically, by assuming the \ac{BS} (and/or user) antennas to be placed in a \ac{ULA} with the common $\lambda/2$ spacing, \cite{boeck24_icml} utilizes a grid  $\mathcal{G}_\mathrm{R}$, whose grid-points lie in the corresponding angular domain $[\pi/2,\pi/2)$. The resulting dictionary $\bm{D}$ then contains steering vectors of the form
\begin{equation}
    \label{eq:steering}
    \bm{a}(\delta) = [1,\operatorname{e}^{-\operatorname{j}\pi \text{sin}(\delta)},\ldots,\operatorname{e}^{-\operatorname{j}\pi (N-1)\text{sin}(\delta)}]^{\operatorname{T}}
\end{equation}
as columns with $\delta \in \mathcal{G}_\mathrm{R}$, and $N$ is the number of antennas. Each $\bm{a}(\delta)$ encodes a specific angle and is associated with one index in the vector $\bm{s}$ in \eqref{eq:s_z}. Moreover, the corresponding entry in $\bm{s}$ represents the associated path loss \cite{boeck24_icml}. Thus, by learning the model parameters $\{\bm{\gamma}_k,\rho_k\}_{k=1}^K$ in \eqref{eq:y_s}-\eqref{eq:z}, we learn a statistical model for the physical channel parameters. Moreover, by considering $\bm{h} = \bm{D}\bm{s}$ together with \eqref{eq:s_z} \& \eqref{eq:z}, we conclude that learning $\{\bm{\gamma}_k,\rho_k\}_{k=1}^K$ implicitly yields a statistical model for the channel $\bm{h}$ itself, i.e.,
\begin{align}
    \label{eq:h_z}
    \bm{h}|k \sim p(\bm{h}|k) = \mathcal{N}_\mathbb{C}(\bm{s};\bm{0},\bm{D}\text{diag}(\bm{\gamma}_k)\bm{D}^{\operatorname{H}}),\\ 
    \label{eq:z2}
    k \sim p(k) = \rho_k.
\end{align}

\section{Reinterpreting the Structured GMM Covariances}
\label{sec:reinterpretation}
As a first step towards jointly estimating parameters and channels, we compare the covariance in \eqref{eq:C_decom} with the learned covariance by \ac{SBGM} for the channel $\bm{h}$ in \eqref{eq:h_z}, i.e., $\bm{D}\text{diag}(\bm{\gamma}_k)\bm{D}^{\operatorname{H}}$. By choosing the number $S_\mathrm{R}$ of grid-points in $\mathcal{G}_\mathrm{R}$ to be $N$ (or $2N$), the dictionary $\bm{D}$ matches in dimension with $\bm{Q}$ in \eqref{eq:C_decom}. The entries $\bm{Q}_{m,n}$ of the \ac{DFT} matrix are $\operatorname{e}^{-\operatorname{j}2\pi (m-1)(n-1)/N}$ (or $\operatorname{e}^{-\operatorname{j}2\pi (m-1)(n-1)/(2N)}$ for the over-sampled version). On the other hand, the entries $\bm{D}_{m,n}$ of the dictionary defined by \eqref{eq:steering} equal $\operatorname{e}^{-\operatorname{j}\pi(m-1)\text{sin}(\delta_n)}$. 

For now, we choose $S_\mathrm{R} = N$ and consider the case $n = 1,\ldots,N/2$ and $n = N/2 + 1,\ldots, N$ separately. By  utilizing $\operatorname{e}^{\operatorname{j}\phi} = \operatorname{e}^{\operatorname{j}\phi-2\pi(m-1)}$ for any $m \in \mathbb{N}$ in the second case, and setting the arguments in $\bm{Q}_{m,n}$ and $\bm{D}_{m,n}$ to be equal, we yield
\begin{align}
\label{eq:circ_theta1}
    \delta^{(\text{circ})}_n &= \arcsin \left(2\frac{n-1}{N}\right)\ \forall\ n = 1,\ldots,\frac{N}{2},\\
    \label{eq:circ_theta2}
    \delta^{(\text{circ})}_n &= \arcsin \left(2\frac{n-1}{N}-2\right)\ \forall\ n = \frac{N}{2}+1,\ldots,N.
\end{align}
Thus, by choosing $\mathcal{G}_\mathrm{R} = \{\delta^{(\text{circ})}_n\}_{n=1}^{S_\mathrm{R} = N}$, applying \ac{SBGM} (cf. \eqref{eq:y_s}-\eqref{eq:z}) leads to a statistical model for $\bm{h}$ (cf. \eqref{eq:h_z} \& \eqref{eq:z2}) which matches the \ac{GMM} investigated in \cite{Fesl2022} with the structured $N \times N$ covariance \eqref{eq:C_decom} and the additional prior knowledge of setting the means $\bm{\mu}_k$ to be zero \cite{boeck2024wcl}. Equivalently, when choosing the over-sampled \ac{DFT} in \eqref{eq:C_decom}, we derive
\begin{align}
\label{eq:toep_theta1}
    \delta^{(\text{toep})}_n &= \arcsin \left(\frac{n-1}{N}\right)\ \forall\ n = 1,\ldots,N,\\
    \label{eq:toep_theta2}
    \delta^{(\text{toep})}_n &= \arcsin \left(\frac{n-1}{N}-1\right)\ \forall\ n = N+1,\ldots,2N.
\end{align}

The work in \cite{boeck24_icml} chooses the grid-points in $\mathcal{G}_\mathrm{R}$ to be equidistant in $[\pi/2,\pi/2)$. However, from \eqref{eq:circ_theta1}-\eqref{eq:toep_theta2}, we conclude that choosing $\mathcal{G}_\mathrm{R}$ in a way that the $\text{sin}(\delta)$-term in \eqref{eq:steering} is uniformly sampled between -1 and 1 leads to \ac{DFT}-based covariance decompositions of the \ac{GMM}-based estimators investigated in, e.g., \cite{Fesl2022}. While choosing $\mathcal{G}_\mathrm{R}$ in this form leads to the same statistical model for $\bm{h}$ as in \cite{Fesl2022}, this reinterpretation offers the key advantage that we have explicit access to a statistical model for the channel parameters \eqref{eq:s_z} \& \eqref{eq:z}. This model can be used for, e.g., simultaneously estimating the channel parameters without computational overhead, making it particularly interesting for \ac{JCAS}. We also show how this model can reduce the estimator's complexity and model parameters.

\section{Joint Parameter and Channel Estimation with the CSGMM}

Since the statistical model in \eqref{eq:s_z} \& \eqref{eq:z} forms a \ac{GMM}, we utilize the \ac{GMM}-based decomposition of the \ac{CME} described in \eqref{eq:gmm_cme} for estimating $\bm{s}$ given a channel observation $\bm{y}$, i.e.,
\begin{equation}
 \label{eq:cme_s}
    \E[\bm{s}|\bm{y}] = \E[\E[\bm{s}|\bm{y},k]|\bm{y}] = \sum_{k=1}^K p(k|\bm{y})\bm{\mu}^{\bm{s}|\bm{y},k}_k
\end{equation}
with
\begin{align}
    \label{eq:mu_s}
    \bm{\mu}_k^{\bm{s}|\bm{y},k} &= \text{diag}(\bm{\gamma}_k)\bm{D}^{\operatorname{H}}\bm{A}^{\operatorname{H}}(\bm{A}\bm{D}\text{diag}(\bm{\gamma}_k)\bm{D}^{\operatorname{H}}\bm{A}^{\operatorname{H}} + \sigma_n^2\operatorname{\mathbf{I}})^{-1}\bm{y}\\
     \label{eq:respo_s}
    p(k|\bm{y}) &= \frac{\rho_k \mathcal{N}_\mathbb{C}(\bm{y};\bm{0},\bm{A}\bm{D}\text{diag}(\bm{\gamma}_k)\bm{D}^{\operatorname{H}}\bm{A}^{\operatorname{H}} + \sigma_n^2\operatorname{\mathbf{I}})}{\sum_{i}\rho_i \mathcal{N}_\mathbb{C}(\bm{y};\bm{0},\bm{A}\bm{D}\text{diag}(\bm{\gamma}_i)\bm{D}^{\operatorname{H}}\bm{A}^{\operatorname{H}} + \sigma_n^2\operatorname{\mathbf{I}})}.
\end{align}
Since $\bm{s}$ encodes the angles of the corresponding channel $\bm{h}$, we can apply the generic DoA estimation scheme from \ac{CS} \cite{stoeckle2015} and identify the peaks in the element-wise squared absolute value $|\E[\bm{s}|\bm{y}]|^2$, which then correspond to our DoA estimates. While $\E[\bm{s}|\bm{y}]$ can be used for parameter estimation in this way, we can also directly derive the \ac{CME} for the corresponding channel $\bm{h}$ by the simple matrix-vector multiplication
\begin{equation}
    \label{eq:h_s_cme}
    \E[\bm{h}|\bm{y}] = \bm{D}\E[\bm{s}|\bm{y}]
\end{equation}
resulting in a joint parameter and channel estimator. 

Equivalent to the \ac{GMM}, training the \ac{CSGMM} can be done in an initial offline phase. Moreover, most of the terms in \eqref{eq:cme_s}-\eqref{eq:respo_s} can also be pre-computed. In fact, the only operations in the online operation for channel estimation are the matrix-vector multiplication in \eqref{eq:mu_s}, calculating the argument of the exponential function in the numerator's Gaussians in \eqref{eq:respo_s}, and \eqref{eq:h_s_cme}.\footnote{Note that $(\bm{A}\bm{D}\text{diag}(\bm{\gamma}_k)\bm{D}^{\operatorname{H}}\bm{A}^{\operatorname{H}} + \sigma_n^2\operatorname{\mathbf{I}})^{-1}$ can be pre-computed for a desired range of noise variances.} We assume $N$ to be the number of antennas, $S_\mathrm{R}$ the number of grid-points (i.e., the dimension of $\bm{s}$), and $K$ the number of \ac{GMM} components. For simplicity and without loss of generality, we assume $\bm{A} = \operatorname{\mathbf{I}}$. The first operation takes $\mathcal{O}(NS_\mathrm{R}K)$, the second takes $\mathcal{O}(N^2K)$, and the third takes $\mathcal{O}(NS)$. Since generally $S_\mathrm{R} > N$, naively implementing the estimator requires an online complexity of $\mathcal{O}(N^2K)$. In the following, we utilize the integrated physics in \eqref{eq:y_s}-\eqref{eq:z} to reduce the online complexity to be linear in $N$. 

Note that each entry in $\E[\bm{s}|\bm{y}]$ represents a distinct angle. Thus, the number of entries in $\E[\bm{s}|\bm{y}]$, which are non-negligible, is not to be expected to be significantly larger than the number of paths existent in the corresponding channel $\bm{h}$. The same argumentation can be applied to $\bm{\gamma}_k$ in \eqref{eq:s_z} \& \eqref{eq:z}. This motivates to introduce a new hyperparameter $P$, which is used to control the number of entries in any $\bm{\gamma}_k$ that has to be considered. We propose to apply the offline operation
\begin{align}
    \label{eq:gamma_P}
    \bm{\gamma}_k \in \mathbb{R}_{+}^{S_\mathrm{R}} &\mapsto \tilde{\bm{\gamma}}_k \in \mathbb{R}_{+}^{P}\\
    \bm{D} \in \mathbb{C}^{N \times S_\mathrm{R}} &\mapsto \tilde{\bm{D}}_k \in \mathbb{C}^{N \times P},
\end{align}
where we extract the $P$ largest values in $\bm{\gamma}_k$, save them in $\tilde{\bm{\gamma}}_k$ and also adapt the dictionary $\bm{D}$ to this transformation by extracting the corresponding $P$ columns from $\bm{D}$ for each component $k$. Subsequently, we utilize the Woodburry identity
\begin{align}
\label{eq:woodburry}
    &(\bm{A}\tilde{\bm{D}}_k\text{diag}(\tilde{\bm{\gamma}}_k)\tilde{\bm{D}}_k^{\operatorname{H}}\bm{A}^{\operatorname{H}} + \sigma_n^2\operatorname{\mathbf{I}})^{-1} = \nonumber\\ 
    &\frac{1}{\sigma_n^2}\left(\operatorname{\mathbf{I}} - \bm{A}\tilde{\bm{D}}_k\left(\text{diag}(\tilde{\bm{\gamma}}_k^{-1}) + \frac{1}{\sigma_n^2}\tilde{\bm{D}}_k^{\operatorname{H}}\bm{A}^{\operatorname{H}}\bm{A}\tilde{\bm{D}}_k\right)^{-1}\tilde{\bm{D}}_k^{\operatorname{H}}\bm{A}^{\operatorname{H}}\right)
\end{align}
resulting in the argument of the exponential function in the numerator's Gaussians in \eqref{eq:respo_s} to have the online complexity of $\mathcal{O}(NKP)$.\footnote{Note that the inverse in \eqref{eq:woodburry} and $\bm{A}\tilde{\bm{D}}_k$ can be pre-computed.} Overall, these modifications result in our proposed method having the online complexity of $\mathcal{O}(NKP)$ for the channel as well as parameter estimation.
\section{Experiments}
\begin{figure}[t]
    \centering
    \includegraphics{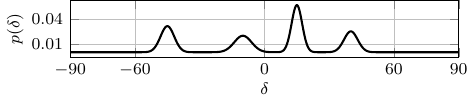}
\vspace{-0.2cm}
\caption{Distribution $p(\delta)$ of the path angle.}
\label{fig:3gpp_gt_angles}
\vspace{-0.6cm}
\end{figure}

\begin{table*}
\begin{minipage}{0.33\textwidth}
\setlength{\tabcolsep}{3pt}
\centering
\captionof{table}{\centering \footnotesize Online complexity - channel estimation.}
\label{tab:1}
\begin{tabular}{cccc}
\toprule
\footnotesize LMMSE & \footnotesize Circ & \footnotesize Toep & \footnotesize CSGMM \\
\midrule
\footnotesize $\mathcal{O}(N^2)$ & \footnotesize $\mathcal{O}(KN\log N)$ & \footnotesize $\mathcal{O}(KN^2)$ & \footnotesize $\mathcal{O}(KNP)$ \\
\bottomrule
\end{tabular}
\label{tab:com_CE}
\end{minipage}
\hfill
\begin{minipage}{0.34\textwidth}
\setlength{\tabcolsep}{3pt}
\centering
\captionof{table}{\centering \footnotesize Online complexity - parameter estimation.}
\label{tab:2}
\begin{tabular}{ccc}
\toprule
\footnotesize SBL & \footnotesize DML & \footnotesize CSGMM \\
\midrule
\footnotesize $\mathcal{O}(T_{\text{iter}}N^3 + S_\mathrm{R})$ & \footnotesize $\mathcal{O}(NS_\mathrm{R})$ & \footnotesize $\mathcal{O}(KNP)$ \\
\bottomrule
\end{tabular}
\label{tab:com_PE}
\end{minipage}
\hfill
\begin{minipage}{0.3\textwidth}
\setlength{\tabcolsep}{3pt}
\centering
\captionof{table}{\centering \footnotesize Number of model parameters.}
\label{tab:3}
\begin{tabular}{ccc}
\toprule
\footnotesize Circ & \footnotesize Toep & \footnotesize CSGMM \\
\midrule
\footnotesize $\mathcal{O}(KN)$ & \footnotesize $\mathcal{O}(KN)$ & \footnotesize $\mathcal{O}(KP)$ \\
\bottomrule
\end{tabular}
\label{tab:number_para}
\end{minipage}
\vspace{-0.45cm}
\end{table*}

\subsection{System and Channel Model}
\label{sec:system_channel}
While our estimator works for the time, frequency, and spatial domain, as well as combinations, we focus on spatial channels, i.e., the \ac{BS} is equipped with $N$ antennas in a \ac{ULA} with $\lambda/2$ spacing. The users have single antennas. When sending pilots, the \ac{BS} receives noisy channel observations 
\begin{equation}
    \label{eq:ch_obs}
    \bm{y} = \bm{h} + \bm{n} \in \mathbb{C}^{N};\ \bm{n} \sim \mathcal{N}_{\mathbb{C}}(\bm{0};\sigma_n^2\operatorname{\mathbf{I}})
\end{equation}
from users within the environment it serves. In the initial offline-phase, the \ac{BS} uses $N_t$ observations $\mathcal{Y} = \{\bm{y}^{(n)}\}_{n=1}^{N_t}$ to train the \ac{CSGMM} according to \cite{boeck2024nips,boeck24_icml}. Note that we do not assume any ground-truth information during training. We assume a constant noise variance throughout the training dataset.~\footnote{Varying noise variances are also possible, cf. \cite{boeck24_icml}.} After training, the \ac{CSGMM} is used for channel and parameter estimation when receiving a newly observed $\bm{y}$. 

The channel model resembles the one in \cite[Appendix H]{boeck24_icml}. Each training and test channel $\bm{h}^{(n)}$ is generated following two steps. First, we draw one angle $\delta^{(n)}$ from the distribution in Fig. \ref{fig:3gpp_gt_angles}. We then draw $\bm{h}^{(n)}$ from $\mathcal{N}_{\mathbb{C}}(\bm{0},\bm{C}_{\delta^{(n)}})$, with 
\begin{equation}
    \label{eq:C_delta}
    \bm{C}_{\delta^{(n)}} = \int_{-\pi}^{\pi} g(\theta;\delta^{(n)})\bm{a}(\theta)\bm{a}(\theta)^{\operatorname{H}}\text{d}\theta,
\end{equation}
$g(\theta;\delta^{(n)})$ is a Laplacian with mean $\delta^{(n)}$ and standard deviation of two degree (cf. \cite{3gppspatial}), and $\bm{a}(\cdot)$ is defined in \eqref{eq:steering}. The distribution in Fig. \ref{fig:3gpp_gt_angles} represents, e.g., four street canyons within a $120^{\circ}$ sector. We apply no other pre-processing.

\subsection{Baselines \& Dictionary}
One baseline for channel estimation is the \ac{LS} estimator $\hat{\bm{h}}_{\text{LS}} = \bm{y}$ (LS). Moreover, we consider the \ac{LMMSE} estimator based on the sample covariance $\bm{S}_y = 1/N_t \sum_n \bm{y}^{(n)}\bm{y}^{(n)\operatorname{H}}$, i.e., $\hat{\bm{h}}_{\text{LMMSE}} = \bm{S}_h\bm{S}_y^{-1}\bm{y}$ with $\bm{S}_h$ being constructed by subtracting the noise covariance from $\bm{S}_y$ and projecting all resulting negative eigenvalues to zero. We also consider the circulant- and Toeplitz-constrained \ac{GMM} estimators (circ \& toep) from \cite{Fesl2022}, where we learn from noisy observations $\mathcal{Y}$ \cite{Fesl2023}. We additionally enforce the means to be zero \cite{boeck2024wcl}. Note that these \acp{GMM} are equivalent to the \ac{CSGMM} with the grid-points in \eqref{eq:circ_theta1}-\eqref{eq:toep_theta2}, and $S_\mathrm{R} = P = N$ and $S_\mathrm{R} = P = 2N$ (cf. \eqref{eq:gamma_P}). As a genie-aided lower-bound baseline, we plot a \ac{LMMSE} estimator (Genie) using the corresponding covariance (cf. \eqref{eq:C_delta}), i.e., $\hat{\bm{h}}_{\text{Genie}} = \bm{C}_{\delta}(\bm{C}_{\delta} + \sigma_n^2\operatorname{\mathbf{I}})^{-1}\bm{y}$.

For parameter estimation, we use the maximum likelihood estimator (DML) that is equivalent to MUSIC for single-DoA estimation \cite{haecker2010}, with $\bm{y}\bm{y}^{\operatorname{H}}$ as subspace-defining covariance \cite{stoeckle2015}. \ac{SBL} has been proposed for parameter estimation with \cite{Shao2022} and without \cite{Gerstoft2016} hyperprior. We consider \ac{SBL} without hyperprior and pre-known noise variance, resulting in the \ac{SBL}-variant from \cite{Wipf2004}. In all simulations, we use the dictionary $\bm{D}$ with the grid $\mathcal{G}_\mathrm{R}$ that contains angles with equidistantly sampled $\sin(\delta)$ term, cf. Section \ref{sec:reinterpretation}.
\subsection{Online Complexity \& Number of Model Parameters}
In Tables \ref{tab:1} and \ref{tab:2}, the online complexity of the channel and parameter estimators are shown, where $T_{\text{iter}}$ is the number of \ac{EM} iterations for \ac{SBL}, which applies the \ac{EM} algorithm online. In the specific case of a single snapshot, DML requires only $\mathcal{O}(NS_\mathrm{R})$ operations. In Table \ref{tab:3}, the number of model parameters is given, which does not scale in $N$ for \ac{CSGMM}. 

\begin{figure}[t]
\hypersetup{hidelinks}
    \centering
    \includegraphics{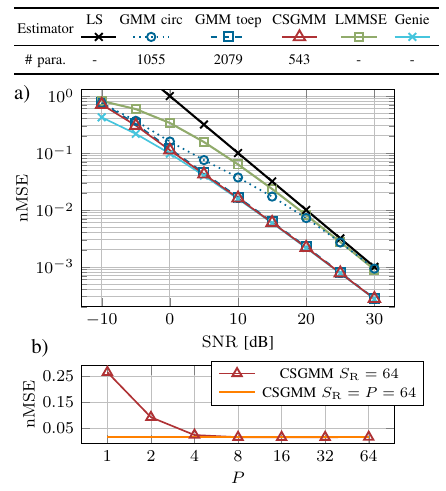}
\vspace{-0.2cm}
\caption{a) $\mathrm{nMSE}$ over $\mathrm{SNR}$ [dB] for $N=32$ with corresponding number of floating point model parameters, b) $\mathrm{nMSE}$ over $P$ for $N=32$, $S_\mathrm{R}=64$, $\mathrm{SNR} = 10$dB.}
\label{fig:nmse_snr}
\vspace{-0.3cm}
\end{figure}
\begin{figure}[t]
\hypersetup{hidelinks}
    \centering
    \includegraphics{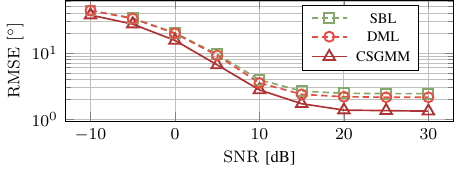}
\vspace{-0.2cm}
\caption{$\mathrm{RMSE}\ [^\circ]$ over $\mathrm{SNR}$ [dB] for $N=16$, $S_\mathrm{R}=256$ and $P=32$.}
\label{fig:rmse_snr}
\vspace{-0.5cm}
\end{figure}
\subsection{Simulation Results}

In Fig. \ref{fig:nmse_snr} a), the \ac{NMSE} $\mathrm{nMSE} = 1/N_{\text{test}}\sum_{j=1}^{N_\text{test}}\|\hat{\bm{h}}^{(j)} - \bm{h}^{(j)}\|^2/N$ with $\hat{\bm{h}}^{(j)}$ being the estimate of the ground-truth channel $\bm{h}^{(j)}$ over the \ac{SNR}, defined as $\mathrm{SNR} = \E[\|\bm{h}\|^2]/(N \sigma_n^2) = 1/\sigma_n^2$ in dB is shown. We choose $N = 32$, $K = 32$, $S_\mathrm{R} = 64$, $P = 16$, $N_t = 20000$, $N_\text{test} = 20000$. Moreover, all trainable models are trained for each \ac{SNR} value separately.\footnote{\ac{CSGMM} also allows varying noise variance in the training set, cf. \cite{boeck24_icml}.} We also present the model's corresponding number of model parameters in the legend. \footnote{Rigorously, \ac{CSGMM} requires additional $KP$ integers, storing the indices for the $P$ largest entries in $\bm{\gamma}_k$, cf. \eqref{eq:gamma_P}, which is negligible.} It can be seen that \ac{CSGMM} as well as \ac{GMM} toep reach the genie-aided lower bound, while the other estimators perform worse. \ac{CSGMM} requires fewer parameters and is computationally cheaper than all baselines. In Fig. \ref{fig:nmse_snr} b), the $\mathrm{nMSE}$ over $P$ for $\mathrm{SNR} = 10$dB is shown. The \ac{CSGMM} performance already converges for $P \geq 4$ and does not require the full $P = S_\mathrm{R} = 64$ as it is realized in \ac{GMM} toep.

One key property of the \ac{CSGMM}-based estimator is that it also directly provides parameter estimates, whose performance is shown in Fig. \ref{fig:rmse_snr}. We plot the \ac{RMSE} $\mathrm{RMSE} = \sqrt{1/N_{\text{test}}|\hat{\delta}^{(j)} - \delta^{(j)}|^2}$ with $\hat{\delta}^{(j)}$ being the estimated angle of the single ground-truth angle $\delta^{(j)}$ existent in the ground-truth channels, cf. Section \ref{sec:system_channel}. We compare DML, \ac{SBL} and \ac{CSGMM} with $S_\mathrm{R} = 256$, $P = 32$, $N = 16$, $N_t = 20000$, $N_\text{test} = 20000$, $K = 32$. \ac{CSGMM} is trained for each \ac{SNR} value separately and outperforms both baselines.

\section{Conclusion \& Future Work}

In this work, we investigated how \ac{SBGM} can be used for joint parameter and channel estimation. We additionally utilized the integrated physics to reduce the complexity and number of model parameters. Next to channel estimation, \acp{CGLM} have been investigated for other physical layer applications such as pilot design \cite{turan2024_pilot}. Similarly, \ac{SBGM} can also be used for these applications, which is part of future work.

\balance
\bibliographystyle{IEEEtran}
\bibliography{references}

\end{document}